# An Integrated Photonic Platform for Rare-Earth Ions in Thin Film Lithium Niobate

Subhojit Dutta[1], Elizabeth A. Goldschmidt[2], Sabyasachi Barik[1], Uday Saha[1], Edo Waks[1*]

[1]*Department of Electrical and Computer Engineering, Institute for Research in Electronics and Applied Physics, and Joint Quantum Institute, University of Maryland, College Park, MD 20742, USA.*
[2]*Department of Physics, University of Illinois at Urbana-Champaign, Urbana, IL 61801*
*\*edowaks@umd.edu*

**Abstract.** Rare-earth ion ensembles doped in single crystals are a promising materials system with widespread applications in optical signal processing, lasing, and quantum information processing. Incorporating rare-earth ions into integrated photonic devices could enable compact lasers and modulators, as well as on-chip optical quantum memories for classical and quantum optical applications. To this end, a thin film single crystalline wafer structure that is compatible with planar fabrication of integrated photonic devices would be highly desirable. However, incorporating rare-earth ions into a thin film form-factor while preserving their optical properties has proven challenging. We demonstrate an integrated photonic platform for rare-earth ions doped in a single crystalline thin film on insulator. The thin film is composed of lithium niobate doped with $Tm^{3+}$. The ions in the thin film exhibit optical lifetimes identical to those measured in bulk crystals. We show narrow spectral holes in a thin film waveguide that require up to 2 orders of magnitude lower power to generate than previously reported bulk waveguides. Our results pave way for scalable on-chip lasers, optical signal processing devices, and integrated optical quantum memories.

**Keywords:** rare-earth ions, thin film lithium niobate, integrated photonics, spectral hole burning, quantum information processing, optical signal processing

Rare-earth ion dopants are solid-state emitters that have found widespread uses in both classical and quantum optics[1–4]. These emitters exhibit stable optical transitions with long lifetimes, making them a useful gain material for lasers and optical amplifiers[4,5]. They also feature narrow homogeneous linewidths,[6,7] which find broad applications in optical signal processing as high finesse filters for laser phase noise suppression,[4,8] and medical imaging[9]. Due to their long coherence times[10], rare-earth ions are also promising candidates for optical quantum memories[11] and qubits[12], which are essential components of quantum networks[13] and distributed quantum computers[14].

The incorporation of rare-earth ions into integrated photonics could enable a new class of active opto-electronic systems with applications in classical and quantum information processing. Integrated photonics combines many optical components on a compact chip, which is essential for scalability and device efficiency. However, incorporating rare-earth ions into integrated photonic devices has proved challenging because these emitters typically reside in bulk crystals[1,4,8], which are not compatible with conventional planar fabrication techniques[15]. One effective solution is to directly pattern integrated photonic devices into the bulk either by ion milling[16–18] or by ion diffusion to form waveguides[1,19]. Hybrid integration[20–22], where nanostructures are evanescently coupled to rare-earth ions in the bulk, provides another promising approach. Implanting rare-earth ions into materials that are compatible with planar fabrication, such as $Er^{3+}$ implanted in silicon nitride[23,24], has been demonstrated, but the homogeneous linewidth is broadened even at cryogenic temperatures[24,25]. A thin film material platform that is single crystalline and preserves the bulk properties of the emitter is therefore critically lacking.

Lithium niobate provides a promising host material for rare-earth ion integrated photonics[1]. Rare-earth doped channel waveguides in lithium niobate have been used to realize lasers[26,27] and quantum memories[19,28,29] spanning a vast range of wavelengths from the infrared to the telecom band. Furthermore, lithium niobate is compatible with commercial wafer scale smart-cut which enables thin single crystalline films on oxide, a lower index material that can support low loss waveguides[30] and also act as a sacrificial layer[31,32]. But the properties of rare-earth ions in smart-cut thin films have not yet been carefully explored.

Here we demonstrate an integrated photonic platform for rare-earth ions in a single crystal thin film which is compatible with scalable top down fabrication. The material stack is composed of smart-cut thin films of $Tm^{3+}$ doped lithium niobate and wafer bonded onto silicon dioxide grown on a undoped lithium niobate substrate. Using this substrate, we pattern waveguides in the $Tm^{3+}$ doped thin film and show strong optical absorption and spectral hole burning through waveguide transmission measurements. We compare the optical properties of the rare-earth ions in the waveguides to those in the bulk and find them to have virtually identical lifetimes and emission spectra, despite significant processing involved in smart-cut and patterning of the thin film. Furthermore, the smart-cut process induces negligible background emission in the substrate. Due to the small cross-sectional area of the waveguides patterned in a thin film, we are able to burn narrow spectral holes with powers that are over two orders of magnitude lower than in previously reported titanium in-diffused waveguides[28]. These results demonstrate the suitability of these emitters for stable and spectrally narrow optical quantum memories, lasers, and filters in a thin film material that is compatible with wafer-scale processing and fabrication.

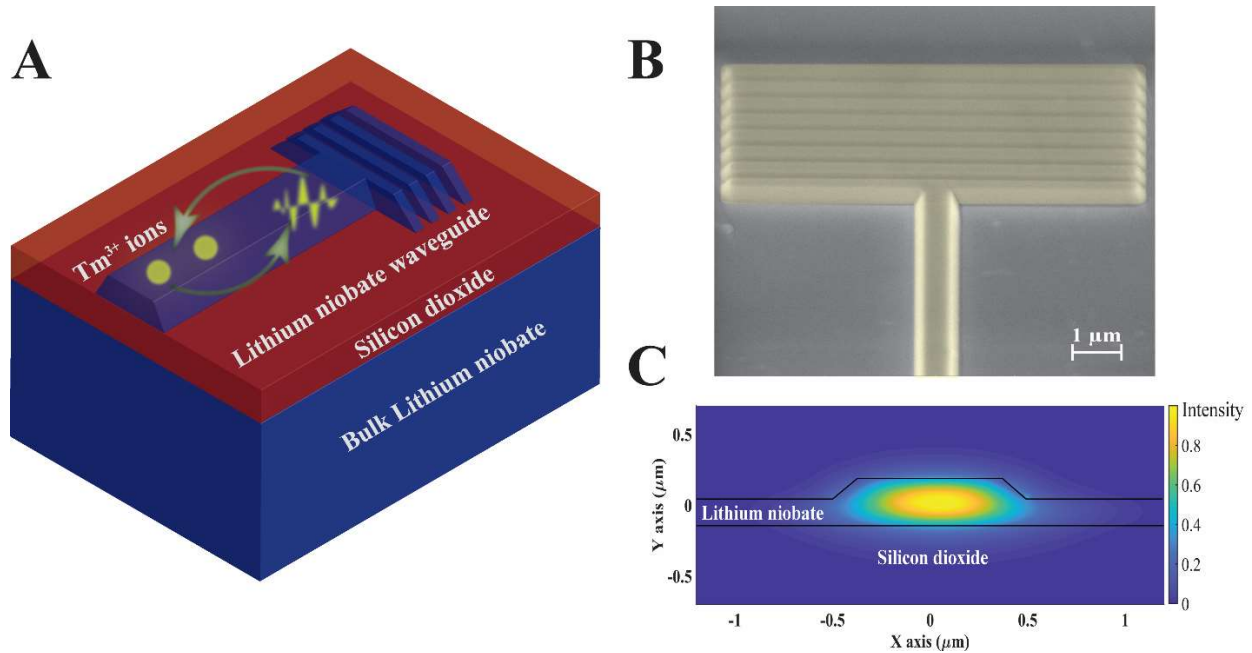

**Figure 1.** (A) Device geometry. Schematic of the cross-section of a waveguide etched in the heterogeneous material stack. (B) Scanning Electron Microscopy image of the waveguide with grating couplers patterned in the $Tm^{3+}$ doped lithium niobate thin film. (C) Finite Difference Time Domain simulation showing the electric field distribution along the waveguide cross-section.

Fig 1(A) shows a schematic of the layer structure of the fabricated device. We begin with a bulk X-cut lithium niobate substrate doped with 0.1% $Tm^{3+}$. Using a commercial smart-cut process (NANOLN), we fabricate a 300 nm layer of doped single crystal lithium niobate wafer bonded to a 2 µm thick layer of silicon dioxide grown on undoped bulk lithium niobate as the substrate material. We etch waveguides into the doped thin film using a two-layer electron beam lithography and dry etching process (see Methods). Fig. 1(B) shows a scanning electron microscope image of the fabricated waveguide along with grating couplers to couple light from the out of plane dimension. The waveguides have smooth sidewalls etched at an angle of 45 degrees. We select the waveguide geometry to be single mode at 794 nm which is the $^3H_6 \rightarrow {}^3H_4$ optical transition for the $Tm^{3+}$ ions. Fig. 1(C) shows a finite difference time domain simulation of the electric field intensity inside the waveguides for the fundamental TE mode at this wavelength. The waveguides are 700 nm wide and constitute only a partial etch of the thin film with an etch depth of 125 nm. Such a design minimizes the interaction of the light with the etched sidewalls resulting in low propagation losses[30]. The optical mode exhibits transverse area of (176 nm X 400 nm) 0.07 µm$^2$, which is over 2 orders of magnitude smaller than titanium in-diffused waveguides[28]. The mode is tightly confined in the thin lithium niobate film where it interacts with the $Tm^{3+}$. The direction of the waveguide must be selected carefully so that the TE mode of the waveguide aligns with the dipole moment of the ions, which are oriented perpendicular to the optical axis (*c* axis) of the lithium niobate substrate [19,28,33]. For comparison, we fabricated waveguides that are both parallel and orthogonal to the *c* axis.

To characterize the sample, we cool it down to 3.6 K in closed cycle cryostat (Attocube attoDRY). We use a confocal microscope with a diffraction limited spatial resolution between the excitation and the collection spots to perform photoluminescence and transmission measurements through the thin film waveguides. To perform broadband measurements, we excite the sample with a supercontinuum source (NKT Photonics) and detect the signal using a spectrometer (Princeton Instruments). For time-resolved measurements and spectral hole burning we use a single frequency laser (M Squared Solstis) modulated by gated acousto-optic modulators (Gooch and Housego) that carve out short pulses with a rise time of 350 ns, and also shift the frequency of the laser (see

Supporting Information Fig. S1). Pulses are detected using a single photon counting module (Excelitas Technologies Inc) with 400 ps temporal resolution.

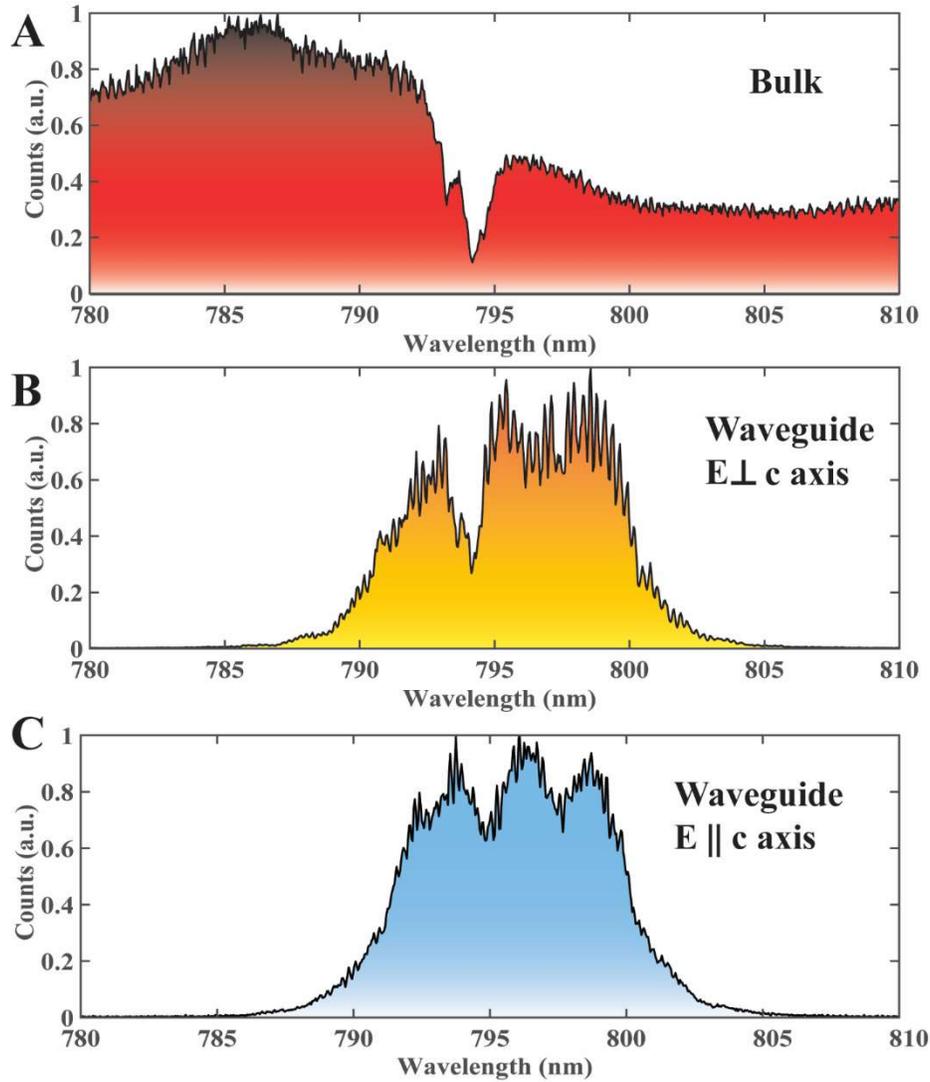

**Figure 2.** (A) Broadband transmission spectrum through a 1 mm length of $Tm^{3+}$ doped bulk lithium niobate crystal with the electric field polarized perpendicular to the crystal *c* axis. (B)-(C) Broadband transmission spectrum filtered down to 10 nm, through a thin film waveguide with the electric field polarized perpendicular (B) and parallel (C) to the crystal *c* axis.

We first characterize the property of the bulk substrate (prior to smart-cut). To perform this measurement, we place the 500 μm-thick bulk substrate on top of a mirror and inject the beam from the top. We excite the sample with the supercontinuum laser and collect the reflected light. In this way the beam performs a double-pass through the bulk and therefore goes through 1 mm of material. Fig. 2(A) shows the spectrum collected through the bulk crystal with the electric field polarization perpendicular to the crystal $c$ axis. The bulk crystal exhibits a strong absorption peak at the wavelength range of 794 nm due to $Tm^{3+}$, which is consistent with past measurements[33]. We observe a peak absorption of 75%, corresponding to an absorption coefficient of 14 cm$^{-1}$ for the ions in the bulk crystal. This number is also consistent with past reported values for bulk $Tm^{3+}$ in lithium niobate with similar doping densities[33]. The sharp absorption peak disappears (not shown) when we orient the polarization parallel to the $c$ axis, as expected due to the local crystal symmetry of the lattice sites occupied by $Tm^{3+}$ ions in lithium niobate.[33].

Next, we probe the thin film by performing waveguide absorption measurements. We excite one end of the waveguide at the grating with a supercontinuum source, filtered to 10 nm using a bandpass interference filter, and collect from the grating coupler at the other end. Fig. 2(B) shows the transmission spectrum through a 775 μm long waveguide with the electric field polarized perpendicular to the crystal axis. The waveguide transmission exhibits the same sharp absorption peak as the bulk at 794 nm. We observe a peak absorption of 66% in the waveguide. This corresponds to an absorption coefficient of 14 cm$^{-1}$, consistent with the bulk value. Fig. 2(C) shows the transmission spectrum through an orthogonal waveguide with the electric field polarized parallel to the crystal $c$ axis. We observe a small dip corresponding to a 30% absorption. This is attributed to the 240 μm bending regions in the long racetrack shaped waveguides where the polarization rotates along the bends and overlaps with the direction perpendicular to the crystal $c$ axis. (see Supporting Information Fig. S2) These measurements indicate that the optical absorption of the ions in the thin film is highly polarized, and that the $Tm^{3+}$ ions retain the bulk absorption properties even after significant nanofabrication steps. The fact that the absorption coefficient is nearly the same for both bulk and waveguides suggests that the waveguide transmission is dominated by the ion absorption, and not by fabrication imperfections in the waveguide that can cause additional undesired loss.

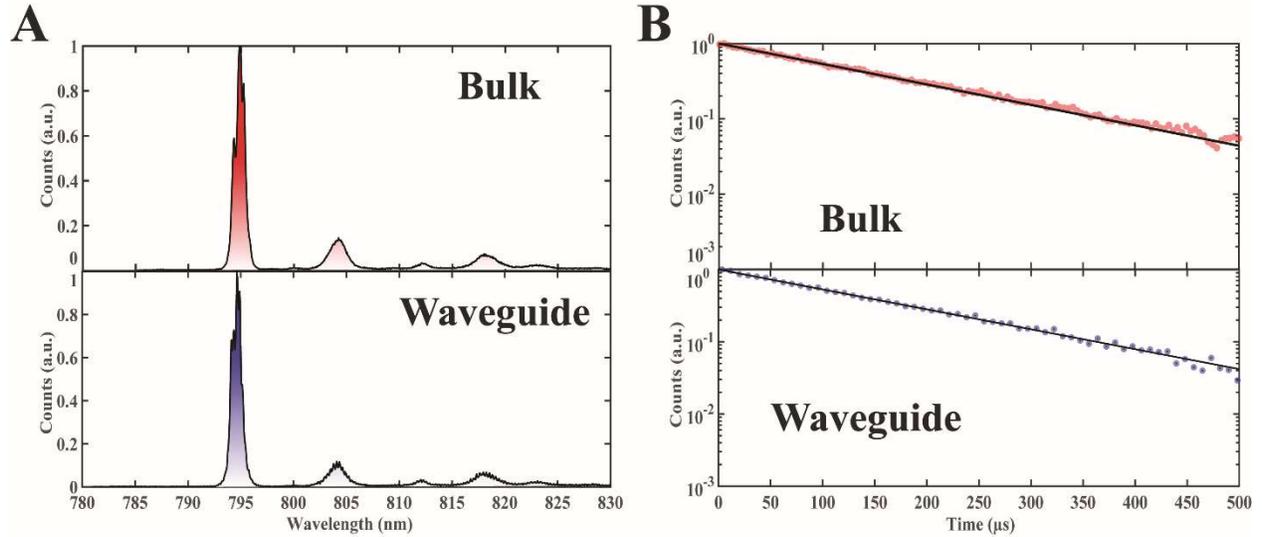

**Figure 3.** (A) Photoluminescence spectrum from bulk crystal and a waveguide in $Tm^{3+}$ doped lithium niobate thin film. (B) Time resolved photoluminescence through a thin film waveguide and bulk crystal respectively. The solid black lines denote a fit to a single exponential decay. The fit reveals lifetimes of 157.4 ± 1.5 μs and 158 ± 2.8 μs in the thin film and bulk respectively.

The inhomogeneous linewidth of the rare-earth ion ensemble is a good measure of the residual lattice strain in the host material. The spectral line shape reflects the local crystal field experienced by the ions. To further probe these properties, we perform photoluminescence spectroscopy of the ions in the bulk sample and the thin film waveguide. We use a narrow tunable laser (M Squared Solstis) to excite the fifth level of the $^3H_4$ excited state multiplet at 773.3925 nm. At 3.6 K the population relaxes to the lowest level of the excited state multiplet by phonon emission on a timescale much shorter than the fluorescence lifetime[33]. Fig 3(A) shows the photoluminescence spectra through the bulk sample and the thin film waveguides. Both spectra reveal similar line shapes for all the transitions from the lowest level of the excited state to the ground state multiplet. The bulk luminescence has a full width at half maximum of 0.7 ± 0.01 nm which is similar to the 0.69 ± 0.01 nm measured for the thin film waveguide. Furthermore, we do not observe appreciable background fluorescence[34] induced by the smart-cut process, which indicates that this process does not generate substantial fluorescence centers that can pollute the $Tm^{3+}$ fluorescence.

In order to ascertain whether the smart-cut process introduces unwanted non-radiative decay, we perform time-resolved photoluminescence measurements. Additional non-radiative decay of in the thin film would lead to a shortened optical lifetime [28,35–37] or non-exponential luminescence decay[36]. We perform time-resolved photoluminescence on the ions in the thin film waveguides and the bulk using a modulated single frequency laser with a center frequency at 773.3925 nm. A gated acousto-optic modulator carves out the excitation laser into short pulses (see Supporting Information Fig. S1). Fig 3(B) shows the time resolved photoluminescence signal from a 15 μm long thin film waveguide and the bulk sample. Both the spectra are fit to a single exponential decay and reveal spontaneous emission lifetimes of 157.4 ± 1.5 μs and 158 ± 2.8 μs respectively, which are equivalent to within the sensitivity of the measurement. We therefore do not observe any additional source of non-radiative decay in the thin film material.

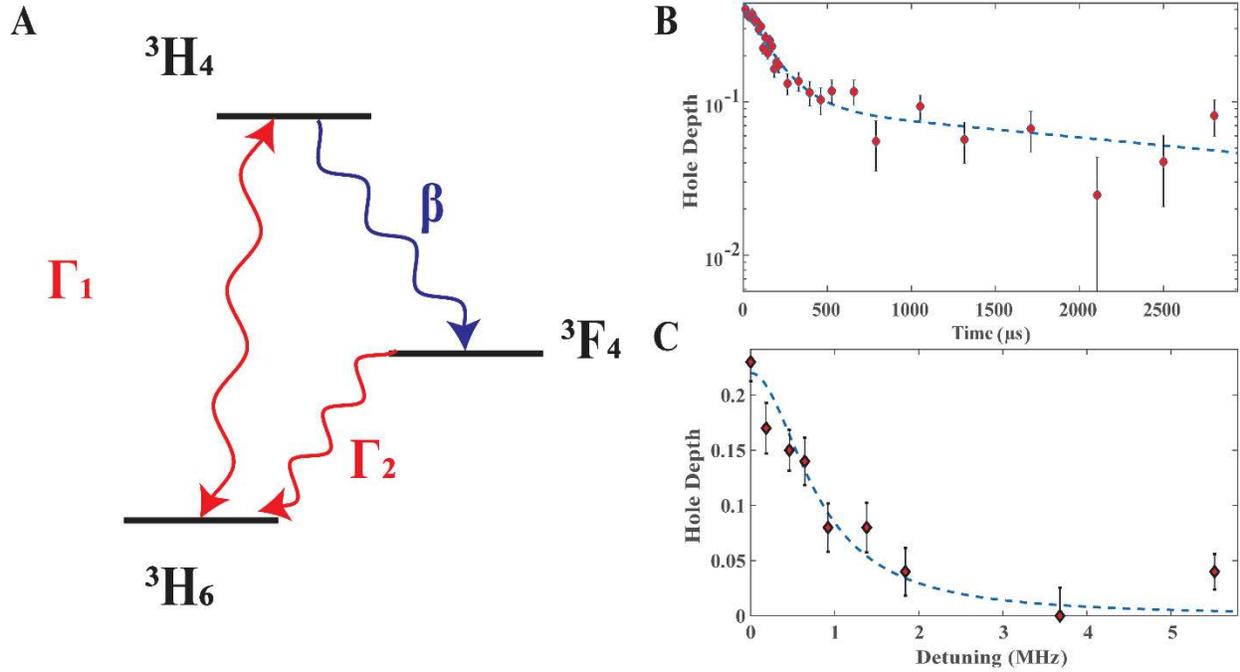

**Figure 4.** (A) Simplified energy level structure of $Tm^{3+}$ ions in a lithium niobate thin film. $\Gamma_1$ and $\Gamma_2$ denote the decay rates from $^3H_4$ and $^3F_4$ to $^3H_6$ respectively. $\beta$ denotes the branching ratio of the transition from $^3H_4$ to $^3F_4$. (B) Time resolved decay of a spectral hole in a thin film waveguide. The dashed line shows a fit to a biexponential decay. (C) Frequency resolved decay of a spectral hole in a thin film waveguide. The dashed line shows a fit to a Lorentzian profile.

One of the key features of a rare-earth ion ensemble is the ability to tailor the inhomogeneously broadened absorption profile by burning spectral holes. Spectral hole burning is a critical step for many applications in both classical and quantum optics[4]. A narrow spectral hole acts as a high quality factor frequency filter for optical signal processing [4,8], and also enables atomic frequency combs that are an essential step for ensemble optical quantum memory protocols [19,38]. $Tm^{3+}$ ions in crystalline lithium niobate have a level structure as shown in Fig. 4(A) which allows for spectral hole burning[19,28,33,39]. Atoms excited to the $^3H_4$ state either decay back to the ground state $^3H_6$ level or to a metastable state $^3F_4$ via two dominant pathways. The shelving state has a long lifetime of several milliseconds. It acts as a population trap before finally decaying back to the ground state. This population trapping allows us to burn a spectral hole in $Tm^{3+}$.

To burn a spectral hole, we split the single frequency laser into two acousto-optic modulators to separately generate a burn pulse and a probe pulse (see Supporting Information Fig. S1). Varying time delays between the burn and the probe pulses maps out the decay of the spectral hole. Fig 4(B) plots the decay of a spectral hole burned at 794 nm in a waveguide patterned on the $Tm^{3+}$ doped thin film. The plot follows a bi-exponential decay with lifetimes of 162 ± 16 µs ($\Gamma_1$) and 3 ± 1.6 ms ($\Gamma_2$) and a branching ratio (β) of 22 ± 6 %. These lifetimes correspond to the spontaneous emission rate of the $^3H_4$ and the decay of the metastable $^3F_4$ respectively, while the branching ratio corresponds to the decay of the $^3H_4$ excited state to the metastable state. The values are consistent with previous measurements performed in bulk crystals[33]. These results suggest that ion sliced materials can be used for atomic frequency combs with similar performance to bulk structures, but in a compact integrated photonic form-factor.

The linewidth of the spectral holes puts a limit on the storage time of the optical quantum memories[19,34,36]. To measure this linewidth, we attenuate the laser burn power and burn a spectral hole to partial depth to limit power broadening. We scan the probe frequency with respect to the burn pulse and map out the decay of the spectral hole. Fig. 4(C) shows the decay of the spectral hole when fit to a Lorentzian profile. The fit shows a linewidth of 1.6 ± 0.4 MHz. The inverse of the spectral hole linewidth puts a lower bound on the optical coherence time at 0.66 ± 0.16 µs. This value is already close to previous reports of a 1.6 µs coherence time of at 3.6 K,[28,33,39] measured using spin-echo. The difference between these measurements is likely due to power broadening and instantaneous spectral diffusion caused by the process of burning a hole and also by instabilities in the laser linewidth in the course of the measurement. Further experiments using two photon echo schemes would provide a more accurate measurement of the actual coherence time of the system.

One advantage of the thin film platform is that it enables waveguides with significantly smaller cross-sectional areas, which significantly reduces the power required to burn spectral holes. Our waveguides have a mode cross sectional area of only 0.07 µm² which is 500 times smaller than that of titanium in-diffused waveguides[28] in bulk lithium niobate. Thus, we expect to be able to burn spectral holes with much less power than that in bulk waveguides. To estimate the power in the waveguide mode we measure the end to end efficiency of the setup from the objective lens that focuses the light onto the input grating coupler, to the collection fiber that collects light from the output grating coupler. We measure this efficiency to be $10^{-6}$. We assume that the efficiencies of the input and output grating coupler are the same, resulting in a coupling efficiency of about $10^{-3}$ for each coupler. Using this efficiency, we calculate that the peak power required to burn a hole to transparency in the waveguide using a 12.5 µs long laser pulse is only 10 nW. This is over two orders of magnitudes lower than previously reported values in bulk waveguides for the same length of the burn pulse[28].

In conclusion, we demonstrate an integrated photonic platform for rare-earth ions in a single crystal thin film which is compatible with scalable planar fabrication. Our technique preserves the desirable optical properties of the rare-earth ions in bulk crystal, paving the way for on-chip lasing, optical signal processing and quantum memory. Lithium niobate also has a strong piezo-electric coefficient that can tune and modulate the ion resonant frequencies to achieve controlled reversible inhomogeneous broadening of narrow absorption lines[28,38] Reducing the waveguide dimensions by going to a thin film substrate enables the creation of strong electric fields at lower voltages and higher bandwidths[40], which would significantly enhance the speed at which the ions can be tuned. Other rare-earth ion species such as $Er^{3+}$ and $Pr^{3+}$ can also be doped into the substrate to enable other wavelength ranges including the telecom band[20,29]. Patterning the substrate into nanophotonic cavities[31] could further enhance light-matter interactions, providing a pathway for spin-photon interfaces[41–43]. Ultimately, our results add tremendous functionality to the emerging field of lithium niobate integrated photonics by providing a compact and versatile optically active material that has broad applications in both classical and quantum optics.

## Methods

**Device Fabrication:** To pattern the thin film waveguide we first grow a 500 nm layer of amorphous silicon which acts as a hard mask. We spin a 300 nm layer of ZEP520A and pattern the samples using a 100 kV electron beam lithography system (Elionix). We deposit 10 nm chromium in an electron beam evaporator and perform metal liftoff to make a negative mask. Using chromium as a mask we first etch the silicon with a standard fluorine chemistry dry etching recipe. Now, with the silicon as a hard mask we transfer the pattern to the thin film lithium niobate using an optimized argon plasma physical etching technique. Finally, we use a 30% solution of KOH at 80° C to remove the remaining silicon. We grow a 2 μm layer of oxide on top of the waveguides to reduce scattering induced losses at the waveguide sidewalls.

## AUTHOR INFORMATION

**Corresponding Author**

edowaks@umd.edu

**Author Contributions**

S. D. and E. W. conceived the experiments. S. D., E. A. G. and E. W. designed the experiments and prepared the manuscript. S. D. conducted the measurements and analyzed the data. S. B. contributed to the optical setup and measurements. U. S. contributed to sample fabrication.

**Funding Sources**

This work was supported by the National Science Foundation (grant number EFMA1741651), the Air Force Office of Scientific Research MURI FA95501910172, the ARL Center for Distributed Quantum Information, and the Physics Frontier Center at the Joint Quantum Institute.

**Acknowledgements**

The authors declare no competing financial interests.

**Supporting Information Available**

We provide supporting information for the waveguide absorption measurements and the optical setup for time resolved photoluminescence and spectral hole burning measurements.

**Table of Content Graphic**

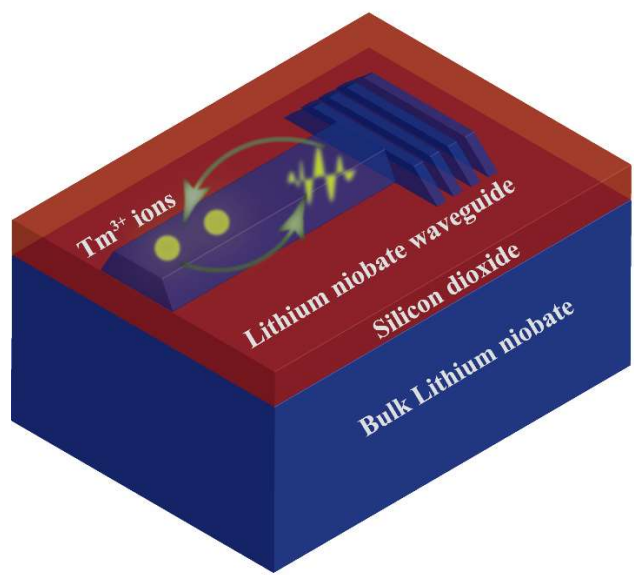
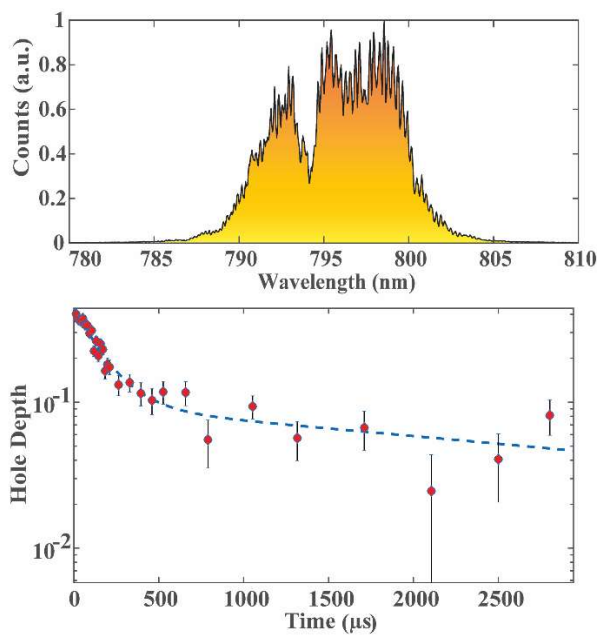

# Supporting Information

# An Integrated Photonic Platform for Rare-Earth Ions in Thin Film Lithium Niobate

**Subhojit Dutta[1], Elizabeth A. Goldschmidt[2], Sabyasachi Barik[1], Uday Saha[1], Edo Waks[1*]**

[1]Department of Electrical and Computer Engineering, Institute for Research in Electronics and Applied Physics, and Joint Quantum Institute, University of Maryland, College Park, MD 20742, USA.
[2] Department of Physics, University of Illinois at Urbana-Champaign, Urbana, IL 61801
*edowaks@umd.edu

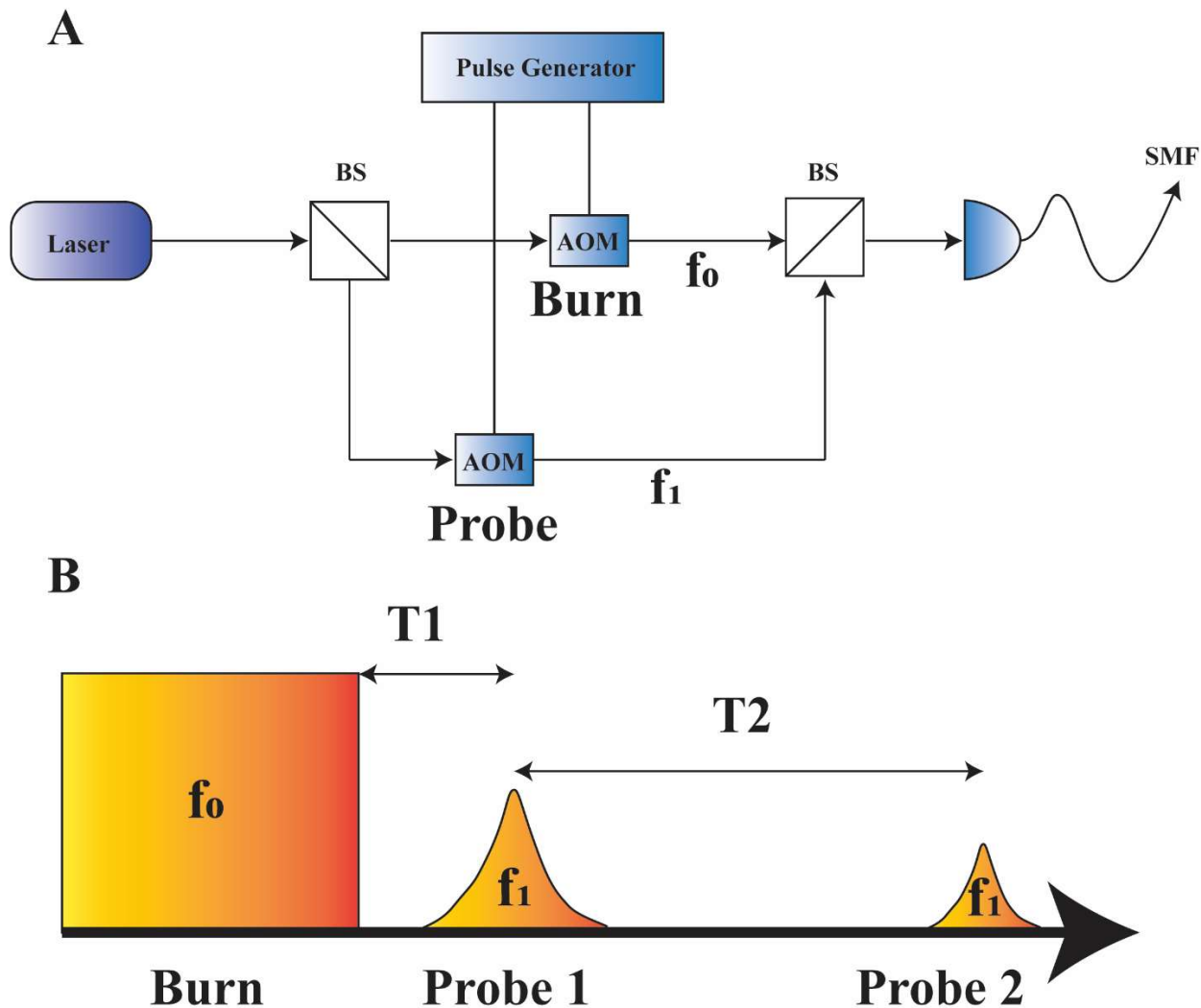

**Figure S1.** (A) Schematic of the setup for generating optical pulses. Laser, Single Frequency MSquared Solistis; BS, Beam Splitter; AOM, Acousto-optic modulator; SMF, Single Mode Fiber; $f_0$, Burn frequency; $f_1$, Probe frequency. (B) Pulse sequence for time and frequency resolved spectral hole burning. T1 denotes the delay between the Burn and Probe1. T2 denotes the delay between Probe1 and Probe2.

**Time Resolved Photoluminescence Experiments:** For time resolved photoluminescence we carve out 500 ms long pulses using one of the AOMs, at 773.3925 nm to excite the fifth level of the $^3H_4$ excited state multiplet in the $Tm^{3+}$ ions. We record the time resolved decay of the photoluminescence signal on a single photon counting module (Excelitas Technologies Inc), spectrally filtered down to a width of 0.1 nm using a spectrometer (Princeton instruments) as a filter.

**Spectral Hole Burning Experiments:** To perform spectral hole burning measurements we use two AOMs to carve out pulses at two independent frequencies, $f_0$ and $f_1$, from the single frequency input laser. We can switch the AOMs by means of an RF switch (Mini circuits) and control pulses generated by an arbitrary function generator (Tektronix). We drive the AOMs using a pair of Voltage Controlled Oscillators (Mini circuits) which can be tuned to independently control the output frequencies $f_0$ and $f_1$.

1) For time resolved spectral hole burning experiments, we use a 12.5 μs long burn pulse with a pulse power of 30 μW before the objective lens, at a frequency $f_0$. We generate a probe with a pulse power of about 0.3 μW before the objective lens. We tune the frequencies of the burn and the probe pulses on resonances by interfering them on a fast photodetector (Thorlabs) and minimizing the observed beat frequency. To perform the experiment, we use a long delay, T2 of about 5 ms between two probe pulses vary the delay T1. The difference in the integrated area between the two probes represents the depth of a spectral hole at a time, T1.

2) For frequency resolved spectral hole burning experiments, we attenuate the burn pulse to about 3 μW before the objective lens. Keeping T1 and T2 fixed we sweep the frequency, $f_1$ of the of the probe with respect to the frequency $f_0$ of the burn pulse and map out the difference in the integrated areas between the two probe pulses which represents the hole depth.

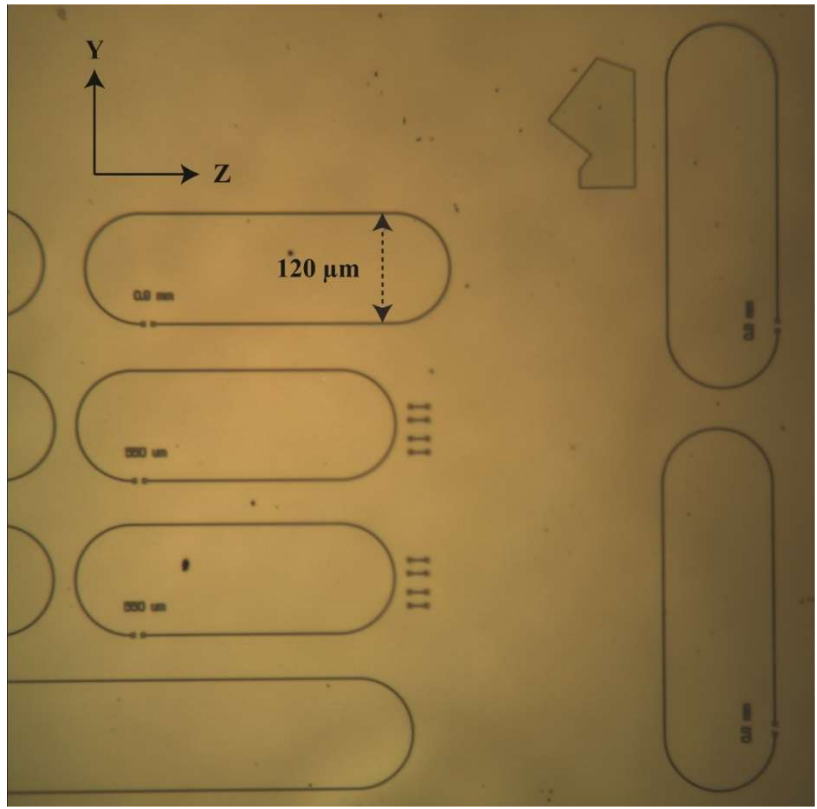

**Figure S2.** Optical microscope image of a representative sample. The crystal *c* axis is oriented along the Z direction. Racetrack shaped waveguides are fabricated orthogonally with respect to the *c* axis of the crystal.

**Waveguide Absorption Experiments:** Figure S2 shows a widefield optical microscope image of racetrack shaped waveguides fabricated in thin film lithium niobate oriented orthogonal and parallel to the crystal axis. Our measurement microscope has a 40 μm wide field of view. To accommodate both the excitation and the collection within the same field of view we fabricate long racetrack shaped waveguides that wrap around. As a result, the polarization of light rotates along the bends and overlaps with the orthogonal direction.